\documentclass[AMA,STIX1COL]{WileyNJD-v2}
\articletype{Research article}%
\received{XX.XX.XXXX}
\revised{YY.YY.YYYY}
\accepted{ZZ.ZZ.ZZZZ}

\usepackage{graphicx}  
\usepackage{booktabs}  
\usepackage{xcolor}

\providecommand{\berndistn}{\mathrm{Bernoulli}}
\providecommand{\betadistn}{\mathrm{Beta}}

\providecommand{\cauchydistn}{\mathrm{Cauchy}}
\providecommand{\normaldistn}{\mathrm{Normal}}

\providecommand{\expodistn}{\mathrm{Exponential}}
\providecommand{\halfnormaldistn}{\mathrm{Halfnormal}}

\providecommand{\dcjs}{D_{\mathrm{CJS}}}

\DeclareMathOperator{\expect}{E}
\DeclareMathOperator{\var}{Var}

\begin{document}

\title{Bayesian random-effects meta-analysis of aggregate data on clinical events}

\author[1]{Christian Röver}
\author[1,2]{Qiong Wu}
\author[3]{Anja Loos}
\author[1,4]{Tim Friede}
\authormark{C.~R\"{o}ver, Q.~Wu, A.~Loos, T.~Friede}

\address[1]{\orgdiv{Department of Medical Statistics}, \orgname{University Medical Center G\"{o}ttingen}, \orgaddress{\state{G\"{o}ttingen}, \country{Germany}}}
\address[2]{\orgdiv{Institute for Medical Biometry and Statistics}, \orgname{University of Marburg}, \orgaddress{\state{Marburg}, \country{Germany}}}
\address[3]{\orgdiv{Global Biostatistics and Epidemiology}, \orgname{Merck KGaA}, \orgaddress{\state{Darmstadt}, \country{Germany}}}
\address[4]{DZHK (German Center for Cardiovascular Research), partner site Lower Saxony, Göttingen, Germany}
\corres{Christian R\"{o}ver, \email{christian.roever@med.uni-goettingen.de}}

\abstract[Summary]{
To investigate intervention effects on rare events, meta-analysis techniques are commonly applied in order to assess the accumulated evidence. When it comes to adverse effects in clinical trials, these are often most adequately handled using survival methods.
A common-effect model that is able to process data in commonly quoted formats in terms of hazard ratios has previously been proposed for this purpose. 
In order to accommodate potential heterogeneity between studies, we have extended the model by Holzhauer to a random-effects approach. The Bayesian model is described in detail, and applications to realistic data sets are discussed along with sensitivity analyses and Monte Carlo simulations to support the conclusions.

}
\keywords{meta-analysis, heterogeneity, prior distribution, hierarchical model}

\jnlcitation{
\cname{\author{C. R\"{o}ver},  \author{Q. Wu}, \author{A. Loos}, and \author{T. Friede}}  (\cyear{2025}), \ctitle{Bayesian random-effects meta-analysis of aggregate data on clinical events}, \cjournal{(submitted for publication)}.
}

\maketitle

\section{Introduction}\label{sec:introduction}
  In order to evaluate risks and benefits of medical treatments, the careful investigation of intended or unintended effects is an integral part of drug development, and meta-analysis methods play a central role here in order to appreciate the entirety of available evidence.
  The assessment of effects on the occurrence of rare events becomes particularly challenging whenever events are so rare that the question arises whether events are observed at all in a relevant group of subjects within a reasonable timeframe.\citep{BradburnEtAl2007}
  Problems of this kind arise commonly in the exploration of adverse events, where  additional challenges arise due to the commonly heterogeneous reporting of events.\citep{UnkelEtAl2019}
  Holzhauer~(2017) proposed a meta-analytic approach to assess effects on the occurrence of medical events within a survival modeling framework and based on commonly reported aggregated event counts as input data.\citep{Holzhauer2017}
  Use of Bayesian methods facilitated the consideration of additional external data via the use of a meta-analytic-predictive (MAP) prior.\citep{SchmidliEtAl2014}
  For the overall (hazard ratio, HR) outcome, a common-effect (CE) model was implemented, while the use of random-effects (RE) models is commonly advocated in meta-analysis, in order to account for (potential) between-study variability (heterogeneity).\cite{HedgesOlkin,HartungKnappSinha,BorensteinEtAl,SchwarzerCarpenterRuecker,SchmidStijnenWhite} In particular with only few studies, the Bayesian framwork offers some advantages over frequestist approaches.\citep{FriedeRoeverWandelNeuenschwander2017a, BenderEtAl2018}
  Here we present an extension of Holzhauer's model, implementing an RE approach. In addtion to the example considered by Holzhauer, namely rosigliatazone in type-2 diabetes, we discuss an application example in oncology, where sample sizes, event rates and numbers of studies are commonly very low, and proper modeling hence is particularly crucial.\citep{Wu2024} These are also distinguishing features between the two examples.

  The remainder of this article is structured as follows. Section~\ref{sec:ma_for_ae} introduces the context, as well as two example data sets that are investigated more closely later on. Section~\ref{sec:methods} explains the common-effect model as well as its random-effects extension in the Bayesian framework. In Section~\ref{sec:exampleAppli}, application of the models to the two example data sets is explained, and the results are compared. Section~\ref{sec:simulStudy} investigates more closely some of the models' operating characteristics using Monte Carlo simulations that are motivated by the example settings. Section~\ref{sec:discussion} eventually closes with a discussion.

\section{Meta-analysis for adverse events}\label{sec:ma_for_ae}
\subsection{Study design and reporting}\label{sec:study_design}
  Studies are commonly designed and powered for an overall treatment effect, often relating to treatment efficacy of some intervention that is in the focus of the investigation. Therefore, they are usually underpowered to infer effects on (rare) adverse events; meta-analyses combining several studies hence promise to yield more conclusive evidence in such a context.
  While information on adverse events is uncontroversially considered important, reporting is commonly rather heterogeneous, and usually only some kind of aggregate count data are available.
  Ideally, one would like to be able to perform a fully-fledged pooled time-to-event (survival) analysis, including accounting for competing events.\cite{CrowtherEtAl2012,BennettEtAl2013,UnkelEtAl2019}
  Such an analysis would require carefully aligned \emph{individual-participant data (IPD)} from all studies;\cite{RileyTierneyStewart} in practice, however, such data are rarely available and one usually needs to settle with aggregate count data.

\begin{figure}
    \centering\makebox{\includegraphics[width=0.30\linewidth]{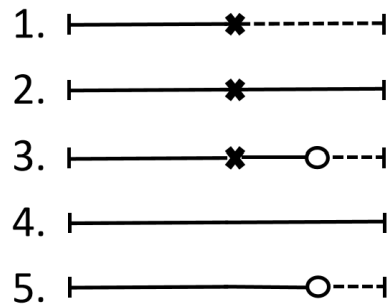}}
    \caption{\label{fig:FiveCat} An illustration of the five different possible patterns showing how fatal or non-fatal events and drop-out may occur over the course of a patient's follow-up time. 
    Follow-up starts on the left, and during the trial duration, adverse events (fatal or non-fatal, $\times$) and drop-out ($\bigcirc$) may or may not occur in different orders. A fatal event or drop-out may terminate the follow-up (dashed lines).
    See Section~\ref{sec:study_design} for more details.}
  \end{figure}

  In the context of adverse events, two issues complicate the analysis. Firstly, as in many time-to-event analyses, the \emph{drop-out} of patients is a problem that needs to be accounted for in the analysis. Often a non-negligible number of patients terminate a study early (for a range of reasons) and hence are not observed for the full planned follow-up time. Secondly, adverse events are commonly analysed as a \emph{composite} of several types of events; some of these may be \emph{fatal}, effectively also terminating the follow-up period for that patient.
  These two problems lead to different types of individual patient timelines that may be summarized by the following five mutually exclusive and collectively exhaustive categories (see Figure~\ref{fig:FiveCat}):\citep{Holzhauer2017}
  \begin{enumerate}
    \item a patient experiences a fatal event of interest during follow-up
    \item a patient experiences a non-fatal event of interest and completes the trial
    \item a patient experiences a non-fatal event of interest and then drops out before completing the trial
    \item a patient completes the study or is still on treatment at the time of analysis without an event of interest (administratively censored)
    \item a patient is lost to follow-up before experiencing any event of interest or completing the trial
  \end{enumerate}
  
  The individual patients' full trajectories of events of interest associated to the clinical study, however,  are usually not available in publications. Instead \emph{aggregate data (AD)}, summaries of IPD data by study arms, are reported. Examples of commonly available arm-level AD include (but are not limited to): the total number of patients ($n_{ij}$), the number of patients who experience medical events of interest ($y_{ij}$), the number of patients who discontinue earlier from the study ($z_{ij}$), the number of patients who experience an event of interest with fatal outcome ($m_{ij}$), and the follow-up duration ($\tau_{ij}$), where $i$~is the trial index and $j$~is the arm index.
  A pragmatic analysis might simply be based on the numbers of events~($y_{ij}$) among the total number of patients~($n_{ij}$), reducing each study's data to a fourfold table and allowing to compute odds ratios (ORs) for these.
  An alternative option might be to also consider the follow-up time~($\tau_{ij}$) and use a Poisson model to derive effects in terms of incidence rate ratios (IRRs).\cite{SeideEtAl2019} Coherent models incorporating all five figures for each study will be introduced in Section~\ref{sec:methods} below.

\subsection{Example data}\label{sec:examples}
  \subsubsection{Rosiglitazone data}\label{sec:RosiData}
    Rosiglitazone (Avandia) is a medication used in the management of type-2 diabetes;\cite{EMA-Avandia} a previous meta-analysis based on a substantial number of 42~studies had suggested a potential association with increased risk of myocardial infarction or cardiovascular mortality.\cite{NissenWolski2007a,Cohen2010,PouwelsGrootheest2012}
    Holzhauer (2017) considered an extended data set of 54~controlled trials with rosiglitazone, including an additional 64~studies providing information on ``historical'' control groups of studies with other drugs conducted over the same period, that had been presented to the Food and Drug Administration (FDA) in~2010.\cite{Holzhauer2017}
    This data set included type-2 diabetes trials with durations from 2~months to 2~years, rosiglitazone doses of~4 or 8~mg, and focused on the risk of \emph{major adverse cardiovacular events (MACE)}. 
    In 15 of the 54 rosiglitazone trials (28\%), no event occurred in any treatment arm, and in an additional 21 of the rosiglitazone trials (39\%), events only occured in one treatment arm. Additionally, in 66\% of trials, a higher proportion of control group than rosiglitazone patients did not complete the trial.  
%
    The actual data (all five figures from Section~\ref{sec:study_design} for each treatment arm) were provided as a supplement to the original publication.\cite{Holzhauer2017} Details on the curation of the data set can be found in Holzhauer (2017).

  \subsubsection{Oncology data}\label{sec:OncoData}
    As a second example, we consider a data set based on an oncological indication, in parctiular featuring a smaller number of studies, and no additional ``historical'' control data.
    In contrast to the previous example, such a scenario is more realistic in oncological applications in general.
    The data set contains results from nine late-stage oncology trials with figures manipulated such that the actual data sources are not identifiable. We are interested in estimating the hazard for a specific safety event under the active treatment by comparing the  occurrence of adverse events in treatment and control arms over various indications and in comparison to different controls. No historical trials are available in this example. The actual data were compiled from publicly available sources (publications, trial registries, etc.) and exposure was read off from published Kaplan-Meier curves.\cite{Wu2024}
    Table \ref{tab:OncoData} lists the relevant aggregate data corresponding to the 9~oncology trials studying the same compound.

    \begin{table}[ht]
      \caption{Aggregate data for the nine oncology example studies,
               each table row here corresponds to a treatment arm.
               Differing indication codes correspond to different types of cancer.
               Treatment~1 is the active treatment of interest (any arm including treatment~1 is considered as the ``treatment arm'').
               For the coding of aggregate data, see also Section~\ref{sec:study_design};
               treatment durations~($\tau_{ij}$) are given in months.
               This data set did not include historical study arms.}\label{tab:OncoData}
      \begin{center}
      \begin{tabular}{cccrrrrr}
      \toprule
            &           &            & \multicolumn{5}{c}{aggregate arm-level data} \\
      \cmidrule(lr){4-8}
      trial~($i$) & indication & treatment & $n_{ij}$ & $y_{ij}$ & $z_{ij}$ & $m_{ij}$ & $\tau_{ij}$ \\
      \midrule
      1           &  1         & 1         & 400      & 5        & 65       & 0        & 30  \\
                  &            & 2         & 370      & 2        & 115      & 0        & 20  \\[1ex]
      2           &  1         & 1         & 680      & 16       & 150      & 0        & 65  \\
                  &            & 3         & 500      & 5        & 90       & 0        & 65  \\[1ex]
      3           &  2         & 1         & 245      & 0        & 220      & 0        & 30  \\
                  &            & 4         & 240      & 1        & 220      & 0        & 30  \\[1ex]
      4           &  2         & 1         & 190      & 0        & 180      & 0        & 15  \\
                  &            & 5         & 175      & 1        & 165      & 0        & 15  \\[1ex]
      5           &  3         & 1         & 350      & 15       & 270      & 0        & 35  \\
                  &            & 6         & 350      & 6        & 330      & 0        & 35  \\[1ex]
      6           &  4         & 1+7       & 440      & 31       & 215      & 0        & 25  \\
                  &            & 8         & 440      & 8        & 280      & 0        & 20  \\[1ex]
      7           &  5         & 1         & 190      & 4        & 180      & 0        & 25  \\
                  &            & 9         & 180      & 1        & 180      & 0        & 25  \\[1ex]
      8           &  5         & 1+10      & 330      & 5        & 180      & 0        & 25  \\
                  &            & 10        & 340      & 6        & 165      & 0        & 25  \\[1ex]
      9           &  6         & 1+11      & 350      & 8        & 120      & 0        & 10  \\
                  &            & 11        & 350      & 5        & 110      & 0        & 10  \\
      \bottomrule
      \end{tabular}
      \end{center}
    \end{table}

    Trials in oncology usually have particular characteristics. Some of these are reflected in the oncology example.Typically, the number of trials investigating a new therapy can be very small. In our case, nine controlled trials were available. Furthermore, trials may be quite heterogeneous in terms of indication (type of cancer) and stage of disease. This is reflected by various indications listed in Table~\ref{tab:OncoData}. Each number represents a different type of cancer; even within the same type of cancer, trials may differ in stage of disease. Also, a drug may be applied as monotherapy or in combination with established treatments (e.g., as an add-on to chemotherapy, radiotherapy, following surgery, etc.), depending on the type of cancer and stage of disease. Therefore, the treatment arm can be defined as the group including the therapy of interest (here coded as~``1''), in some cases in combination with other therapies. Depending on the mode of action, a new investigational treatment may be considered suitable in several oncological indications. Hence, although the treatment arms of all trials contain the same compound, established treatment regimen may differ over the indications including treatment in the control groups of different trials.
    
    There are no historical trials in this oncology data example, because it can be difficult to find historical data during clinical development, especially when undertaken in populations with very specific selective characteristics (e.g. newly detected biomarkers). Historical data may become available over time and could also be available from the beginning when several investigational products are examined in similar populations.
    
    Since we focus on safety events in this data example, the difference in data analysis for the purpose of efficacy and safety leads to some caveats regarding our aggregate data. In many cases, disposition of patients is presented for the full analysis set (FAS) including all patients randomized, whereas the safety results are based on the safety analysis set (SAF) commonly defined as  patients who received at least one dose of treatment. In most cases, FAS and SAF are the same, but there may be situations with substantial discrepancies (e.g., a required pre-therapy may not have been completed and a patient did not continue to the main part of the trial). Small discrepancies occur when patients are randomized but never received any treatment or when patients are not treated according to the randomized treatment by error. Consequently, the total number of patients in a certain arm~($n_{ij}$) and the number of patients who experienced adverse events of interest in a certain arm~($y_{ij}$) are based on SAF in our example, but the number of early discontinuations due to competing events~($z_{ij}$) is based on FAS, leading to a little inconsistency that one should keep in mind. What might be considered a competing event is disease specific and is likely to be very different for the examples in oncology and diabeetes considered here.

\begin{figure}[!htb]
        {\centerline{\includegraphics[width=0.80\linewidth]{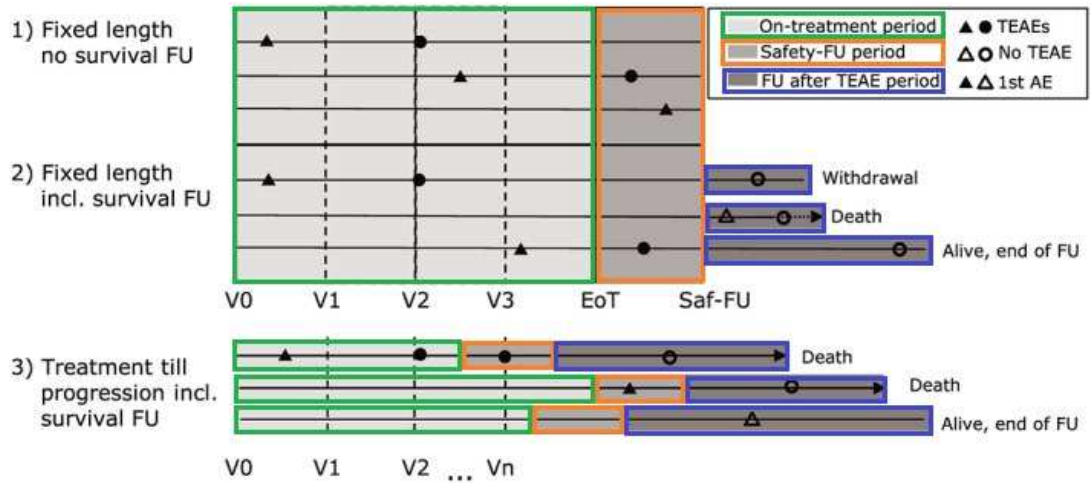}}}
        \caption[Typical scenarios for adverse event follow‐up in clinical trials]{Typical scenarios for adverse event follow‐up in clinical trials (adopted from Figure~1 in Unkel \emph{et~al.}(2019)\cite{UnkelEtAl2019}). The abbreviations are: AE, adverse event; TEAE, treatment-emergent adverse event; FU, follow-up; Saf-FU, safety follow-up; EoT, end of treatment; V$0$, visit at trial onset; V$1$,\ldots,V$n$, visits during treatment. First occurrences of AEs are marked by triangles.}
        \label{fig:UnkelFig1}
    \end{figure}

    Unkel \emph{et~al.} (2019)\cite{UnkelEtAl2019} illustrate the safety follow-up time and efficacy follow-up time in different situations as in Figure~\ref{fig:UnkelFig1}. In earlier stages of disease (e.g. with treatment of curative intent),  treatment protocols may be limited to a fixed number of cycles and patients are followed up for a defined period of time for safety. Information on disease recurrence or survival may continue much longer, which is shown in~2) and~3) of Figure~\ref{fig:UnkelFig1}. In clinical trials of late-stage solid tumors, as in our data example, patients frequently receive treatment until progression of disease, unacceptable toxicity, or withdrawal from treatment. As a result, time on treatment may vary considerably between treatment arms when a treatment is efficacious in prolonging the time to disease progression. In addition, a patient may withdraw from treatment due to toxicity and intolerabilities. Differential time on treatment duration is also common within the same treatment arm, even in the situation of a fixed number of treatment cycles. Upon progression of disease, patients are frequently referred to subsequent anti-cancer treatment with an own toxicity profile or may participate in subsequent clinical trials. In most cases, follow-up for safety stops within a certain time frame after end of treatment or start of new anti-cancer treatment in order to avoid a dilution of the safety profile of the new investigational product. Follow-up for survival usually continues after end of treatment as this does not require regular physical visits at the clinical trial site but might still be difficult to achieve from a practical point of view if for instance patients switch into other studies. Based on the above argument, the maximum duration~$\tau$ in our example should be the varying safety follow-up time, graphically corresponding to the green plus orange part in~3) of Figure~\ref{fig:UnkelFig1}. However, the maximum safety follow-up time is not directly available in the publications, but can be read off as part of result presentation on progression-free survival time (i.e., from Kaplan-Meier curves). The difference between these two are considered to be reasonably small in our example data set.
    
  \subsection{Modelling medical event data}
    The meta-analysis of event count data may be approached in a number of different ways, ranging from simpler or more pragmatic options to more sophisticated models that may be able to better capture specific situations.\cite{DeeksAltman2001,HerbisonEtAl2015,SeideEtAl2019}
    In a context closely related to the rosiglitazone data example from  Section~\ref{sec:RosiData}, Nissen and Wolski (2007) reported on a meta-analysis of adverse effects based on the numbers of events among the total of patients ($y_{ij}$ and~$n_{ij}$ in the terminology from Section~\ref{sec:study_design}).\cite{NissenWolski2007a,NissenWolski2007b}
    An additional complication here was that a substantial fraction of studies had reported no event in one or both treatment arms, challenging the assumptions underlying some analysis models.\cite{SweetingSuttonLambert2004,BradburnEtAl2007} Furthermore, the original analysis was also based on a common-effect model, which generated subsequent discussions of appropriate modelling approaches in this context.\cite{DiamondBaxKaul2007,RueckerSchumacher2008}
    Unkel \emph{et~al.} (2019) later approached the analysis of adverse events from the \emph{estimands} perspective and provided recommendations for the reporting and (meta-) analysis.\cite{UnkelEtAl2019} 

    Holzhauer (2017)\cite{Holzhauer2017} then proposed a coherent Bayesian approach for the kind of aggregate data outlined in Section~\ref{sec:study_design} (and Figure~\ref{fig:FiveCat}), accounting for different event patterns, differing follow-up times, etc., and using the rosiglitazone example from Section~\ref{sec:RosiData} for illustration. This more sophisticated model included assumptions on the (exponential) distribution of waiting times, and in addition allowed for the consideration of ``historical'' control data. The eventual treatment effect, the hazard ratio (HR), was considered in terms of a common effect. This model is described in more detail in the following section and will be generalized subsequently.

\section{Methods}\label{sec:methods}
  \subsection{The common-effect model by Holzhauer}\label{sec:FeModel}
    Suppose the main meta-analysis comprises $I$~randomized controlled trials ($i=1,\ldots,I$). Each trial contains two arms with $j=0$ denoting the control arm and $j=1$ the treatment arm. Historical control group AD may be available from $H$~historical trials ($i=I+1,\ldots,I+H$).
    As outlined in Section~\ref{sec:study_design}, the AD are summarized by sample sizes~$n_{ij}$, event counts~$y_{ij}$ and~$m_{ij}$, numbers of drop-outs~$z_{ij}$ and follow-up duration~$\tau_{ij}$.
    Indices here run from $i=1,\ldots,I+H$, where $j\in\{0,1\}$ if~$i\leq I$ (the two-armed trials), and $j=0$ for $i>I$ (the ``historical'' single-arm trials).

    We then assume exponential distributions for a patient's time lag until an event or until censoring, and a Bernoulli distribution on whether a given event is fatal or not. More specifically,
    for the $k$th patient ($k=1,\ldots,n_{ij}$) in arm~$j$ of the $i$th study, we assume the underlying data generating mechanisms as follows:
    \begin{eqnarray}
      X_{ijk}|\lambda_{ij} & \sim & \expodistn(\lambda_{ij}) \mbox{,} \label{eqn:ExpoAssumption1}\\
      C_{ijk}|\mu_{ij}     & \sim & \expodistn(\mu_{ij}) \mbox{,} \label{eqn:ExpoAssumption2}\\
      M_{ijk}|q_{j}        & \sim & \berndistn(q_{j}) \mbox{.} \label{eqn:BernAssumption}
    \end{eqnarray}
    Here, 
    $X_{ijk}$ refers to time to first event of interest, $C_{ijk}$ means the drop-out time and $M_{ijk}$ indicates whether an event of interest is fatal given that it has occurred. Note that $C_{ijk}$ is not necessarily non-administrative censoring, it could be competing risk especially in the oncology example, such as progression of disease or death due to causes other than the event of interest. For any patient in the analysis, we assume  $X_{ijk}$, $C_{ijk}$ and $M_{ijk}$ are independent. In addition, we assume independence between trials. 

    Let $\boldsymbol{w_{ij}}:=(w_{ij}^{(1)},w_{ij}^{(2)},w_{ij}^{(3)},w_{ij}^{(4)},w_{ij}^{(5)})^T$ denote the numbers of patients observed in each category of patient timelines introduced in Section~\ref{sec:study_design}. This random vector follows a multinomial distribution 
    with event probabilities~$p_{ij}^{(1)},\ldots,p_{ij}^{(5)}$ (and $\sum_{s=1}^{5}p_{ij}^{(s)}=1$).
    Given the distributional assumptions made on event time~(\ref{eqn:ExpoAssumption1}), drop-out time~(\ref{eqn:ExpoAssumption2}) and fatal event occurrence~(\ref{eqn:BernAssumption}), the probabilities $p_{ij}^{(1)},...,p_{ij}^{(5)}$ then result as follows:
    \begin{eqnarray}
        p_{ij}^{(1)} &=& q_j\frac{\lambda_{ij}}{\lambda_{ij}+\mu_{ij}}\Bigl(1-\exp\bigl(-(\lambda_{ij}+\mu_{ij})\tau_i\bigr)\Bigr) \mbox{,}\\ 
        p_{ij}^{(2)} &=&(1-q_j)\bigl(1-\exp(-\lambda_{ij}\tau_i)\bigr) \exp(-\mu_{ij}\tau_i) \mbox{,}\\
        p_{ij}^{(3)} &=&(1-q_j) \biggl( \frac{\lambda_{ij}+\mu_{ij}\exp\bigl(-(\lambda_{ij}+\mu_{ij})\tau_i\bigr)}{\lambda_{ij}+\mu_{ij}} - \exp(-\mu_{ij}\tau_i) \biggr) \mbox{,}\\
        p_{ij}^{(4)} &=& \exp\bigl(-(\lambda_{ij}+\mu_{ij})\tau_i\bigr) \mbox{, and}\\
        p_{ij}^{(5)} &=& \frac{\mu_{ij}}{\lambda_{ij}+\mu_{ij}}\Bigl(1-\exp\bigl(-(\lambda_{ij}+\mu_{ij})\tau_i\bigr)\Bigr) \mbox{.}
    \end{eqnarray}
    In most cases, we only know $w_{ij}^{(1)}$ from the publications, because it corresponds to the $m_{ij}$ in AD, i.e. the number of patients with fatal event of interest. The sizes of the other four categories are not directly available from publications. What we know is that $w_{ij}^{(3)}$ lies between $r_1=\text{max}(0,y_{ij}+z_{ij}-m_{ij}-n_{ij})$ and $r_2=\text{min}(y_{ij}-m_{ij},z_{ij}-m_{ij})$. Given a particular value of $w_{ij}^{(3)}$, we can uniquely determine the values of $w_{ij}^{(2)}$, $w_{ij}^{(4)}$ and $w_{ij}^{(5)}$ as
    \begin{eqnarray}
      w_{ij}^{(2)} &=& y_{ij}-w_{ij}^{(1)}-w_{ij}^{(3)}\mbox{,}\\
      w_{ij}^{(5)} &=& z_{ij}-w_{ij}^{(1)}-w_{ij}^{(3)}\mbox{,}\\
      w_{ij}^{(4)} &=& n_{ij}-w_{ij}^{(1)}-w_{ij}^{(2)}-w_{ij}^{(3)}-w_{ij}^{(5)}\mbox{.} 
    \end{eqnarray}
    Therefore, the likelihood can be calculated by summing over the possible values of $w_{ij}^{(3)}$:
    \begin{equation} \mathcal{L}(p_{ij}^{(1)},...,p_{ij}^{(5)}\mid AD)=n_{ij}!\frac{p_{ij}^{(1)^{m_{ij}}}}{m_{ij}!}\sum_{r=r_1}^{r_2}\frac{p_{ij}^{(2)^{y_{ij}-m_{ij}-r}}}{(y_{ij}-m_{ij}-r)!}\frac{p_{ij}^{(3)^r}}{r!}\frac{p_{ij}^{(4)^{n_{ij}-y_{ij}-z_{ij}+m_{ij}+r}}}{(n_{ij}-y_{ij}-z_{ij}+m_{ij}+r)!}\frac{p_{ij}^{(5)^{z_{ij}-m_{ij}-r}}}{(z_{ij}-m_{ij}-r)!} \end{equation}
%
%
    With this likelihood based on AD given in terms of~$w_{ij}^{(1)}$ , we may proceed to specify the hierarchical model for meta-analysis as follows:
    \begin{eqnarray}
        \log(\lambda_{i0})\mid\nu_{1}, \sigma_{1} & \sim & \normaldistn(\nu_1, \sigma_1^2),  \\
        \log(\mu_{i0})\mid\nu_{2}, \sigma_{2} & \sim & \normaldistn(\nu_2, \sigma_2^2), \\
        \log(\mu_{i1})\mid\nu_{3}, \sigma_{3} & \sim & \normaldistn(\nu_3, \sigma_3^2), \\
        \log(\lambda_{i1}) \mid \lambda_{i0}, \varphi & = & \log(\lambda_{i0}) + \varphi \mbox{,} \label{eqn:FE}
    \end{eqnarray}
    where $\varphi$~is the logarithmic \emph{hazard ratio (HR)}, which is assumed to be common across trials. The (trial-specific) random effects $\log(\lambda_{i0})$, $\log(\mu_{i0})$ and $\log(\mu_{i1})$ are assumed to jointly follow normal distributions with unknown means and variances that are treated as hyperparameters.
    
    Within a Bayesian modeling framework, prior probability distributions need to be specified, reflecting the \emph{a~priori information} on any unknown parameters that is supposed to be reflected in the inference.
    In case of no specific information on the exponential rate parameters~$\lambda_{ij}$ and~$\mu_{ij}$, vague priors may be used for location ($\nu_1$, $\nu_2$, $\nu_3$) and scale parameters ($\sigma_1$, $\sigma_2$, $\sigma_3$), for example, $\normaldistn(0, 100^2)$ priors for the location, and $\halfnormaldistn(100)$ distributions for the scale.\cite{Roever2020,RoeverEtAl2021} A vague specification for the Bernoulli probabilities~$q_j$ may be given by Jeffreys' noninformative prior, a  $\betadistn(0.5, 0.5)$ distribution.\cite{BDA3rd}

    In the presence of external information that should be incorporated in the analysis, the informativeness of priors may be adapted accordingly. For example, in the rosiglitazone data example, there is some information from previous observational data that may be summarized into weakly informative priors on the hyperparameters. Holzhauer (2017) set the prior for the mean and scale of $\log(\lambda_{i0})$ as $\normaldistn(-4.27, \log(10)^2)$ and $\halfnormaldistn(\log(10))$.
    The prior for the mean and scale of $\log(\mu_{i0})$ and $\log(\mu_{i1})$ was $\normaldistn(\log(0.22), \log(10)^2)$ and $\halfnormaldistn(\log(10))$.\cite{Holzhauer2017}
    
    When there are historical trials available, we may borrow information from these trials to form informative priors on the parameters of the main meta-analysis. One way to utilize historical information is using a robust \emph{meta-analytic-predictive (MAP)} prior.\cite{SchmidliEtAl2014,WeberEtAl2021}
    The key idea is to form priors on the parameters of the main trials by combining the predictive distribution for the control group parameters resulting from fitting the hierarchical meta-analysis model to the historical data with a weakly informative mixture component.
    This can be implemented in two ways that differ in whether borrowing of information between trials in the main meta-analysis is allowed.
    The first option is to use a mixture of the resulting posterior for the hyperparameters ($\nu_1$, $\sigma_1$, $\nu_2$, $\sigma_2$, $\nu_3$, $\sigma_3$) and a weakly informative component as the prior for the hyperparameters in the main meta-analysis. Since the hyperparameters are common to all trials, such a specification allows for borrowing of strength among the trials in the main meta-analysis, that is, the estimates of the trial-level nuisance parameters influence each other. We therefore denote such models as ``non-stratified'' models. 
    The second option is to specify a mixture of a weakly informative component and the marginal MAP for trial-level parameters ($\log(\lambda_{i0})$, $\log(\mu_{i0})$ and $\log(\mu_{i1})$) from the meta-analysis of historical data as priors for the trial-level parameters in the main meta-analysis. Such specification results in a ``stratified'' model without borrowing of strength among the trials in the main meta-analysis. 
    For exploratory purposes, we try different weights (50\%, 80\% and 100\%) given to the informative MAP when mixing with a weakly informative component.
    
    For the log-HR~$\varphi$, we propose to employ either a $\cauchydistn(0.0, 2.5)$ distribution as weakly informative prior\cite{GelmanEtAl2008} or a $\cauchydistn(0.0, 0.37)$, which puts 90\% of the prior probability mass within $[-\log(10), \log(10)]$ reflecting the belief that extremely large drug effects with odds ratios above 10 or below 0.1 are rare. \cite{HamraEtAl2013,Greenland2000}

  \subsection{The proposed random-effects model}\label{sec:ReModel}
    The above model implements random effects for the time-to-event parameters ($\log(\lambda_{ij})$, $\log(\mu_{ij})$) and common parameters for the two proportions of fatal events ($q_1$, $q_2$). The effect of primary interest, the log-HR~$\varphi$, is also treated as a \emph{common effect (CE)}.
    In the meta-analysis context, it is very common to consider (potential) between-trial heterogeneity in terms of \emph{random effects (RE)}, as was also done for the time-to-event parameters;
    neglecting heterogeneity may in particular lead to ``na\"{i}ve'' pooling of evidence and overconfidence in results.\cite{BorensteinEtAl2010,KontopantelisSpringateReeves2013}
    The aim here is to generalize Holzhauer's original approach to also account for (potential) heterogeneity in the overall HR\@.

    To generalize Holzhauer's model, we may adopt most of the original specifications, except for the model for the log-HR~(\ref{eqn:FE}), which is then defined as
    \begin{eqnarray}
        \log(\lambda_{i1}) \mid \lambda_{i0}, \varphi_i &=& \log(\lambda_{i0}) + \varphi_i, \\
        \varphi_i \mid \varphi, \eta &\sim& \normaldistn(\varphi, \eta^2), \label{eqn:RE}
    \end{eqnarray}
    where $\varphi_i$ now is the \emph{trial-specific} log-HR for the $i$th~trial. The overall mean effect is denoted by~$\varphi$, and $\eta$ reflects the between-trial heterogeneity. In the special case of~$\eta=0$ (homogeneity), the model again reduces to the CE~model.

    For the overall log-HR~$\varphi$, it may make sense to adopt the same priors ($\cauchydistn(0.0, 2.5)$ or $\cauchydistn(0.0, 0.37)$) as in Section~\ref{sec:FeModel}. 
    A $\halfnormaldistn(100)$ distribution may be used as a vague prior on the scale~$\eta$, essentially as a conservative specification mimicking an (improper) uniform prior.\citep{Roever2020}
    Otherwise, a $\halfnormaldistn(0.5)$ distribution  is used as a weakly informative prior on~$\eta$, especially when there is a small number of trials in the main meta-analysis.\cite{RoeverEtAl2021}

  \subsection{Implementation of the models}
    The Bayesian analyses are carried out using Markov chain Monte Carlo (MCMC) methods.\cite{MCMCinPractice}
    The CE~model was originally implemented using the the \texttt{proc~MCMC} procedure within the \textsf{SAS}~software, while a \textsf{Stan} implementation has also been used for simulations.\citep{Holzhauer2017,HolzhauerSupplement}
    The random-effects extension was now implemented in \textsf{R} (version~4.3.2) and \textsf{Stan},\cite{HoffmanGelman2014} as well as the \texttt{rstan}\cite{R:rstan} (version~2.32.5) and \texttt{RBesT}\cite{WeberEtAl2021}(version~1.7-3)~packages.
    Derivation of MAP priors was implemented using the \texttt{mclust}, \texttt{teigen} and \texttt{RBesT} functions.\cite{Wu2024}
    \textsf{Stan} implements a \emph{Hamiltonian Monte Carlo} sampler; 
    each time, $5$~chains were run in parallel,
    to ensure proper burn-in, the initial $10\,000$~samples of each chain were discarded, 
    and to reduce autocorrelations, only every $5$th sample of each chain was kept.
    Simulations eventually lasted approximately $10\,000$ samples for each iteration, and convergence was checked using the rank-based~$\hat{R}$.\citep{VehtariEtAl2021}
    Prior sensitivity is quantified using the $\dcjs$~metric as implemented in the \texttt{priorsense} \textsf{R}~package, where smaller values indicate less sensitivity of results to changes in the corresponding parameter's prior specification.\citep{KallioinenEtAl2024,R:priorsense}
    
    The results from the re-implementation were then cross-checked against the originally published estimates.\cite{Holzhauer2017} The resulting estimates were mostly in good agreement, except for some slight discrepancies that may most likely be attributed to poor mixing of MCMC~chains in the \textsf{SAS}~implementation.\cite{Wu2024}

\section{Example applications}\label{sec:exampleAppli}
  Returning to the example data introduced in Section \ref{sec:examples} we illustrate the application of the models and discuss their differences. First the rosiglitazone data are considered.

  \subsection{Rosiglitazone data}\label{sec:RosiAppli}
\begin{figure}[!htb]
        {\centerline{\includegraphics[scale=0.6]{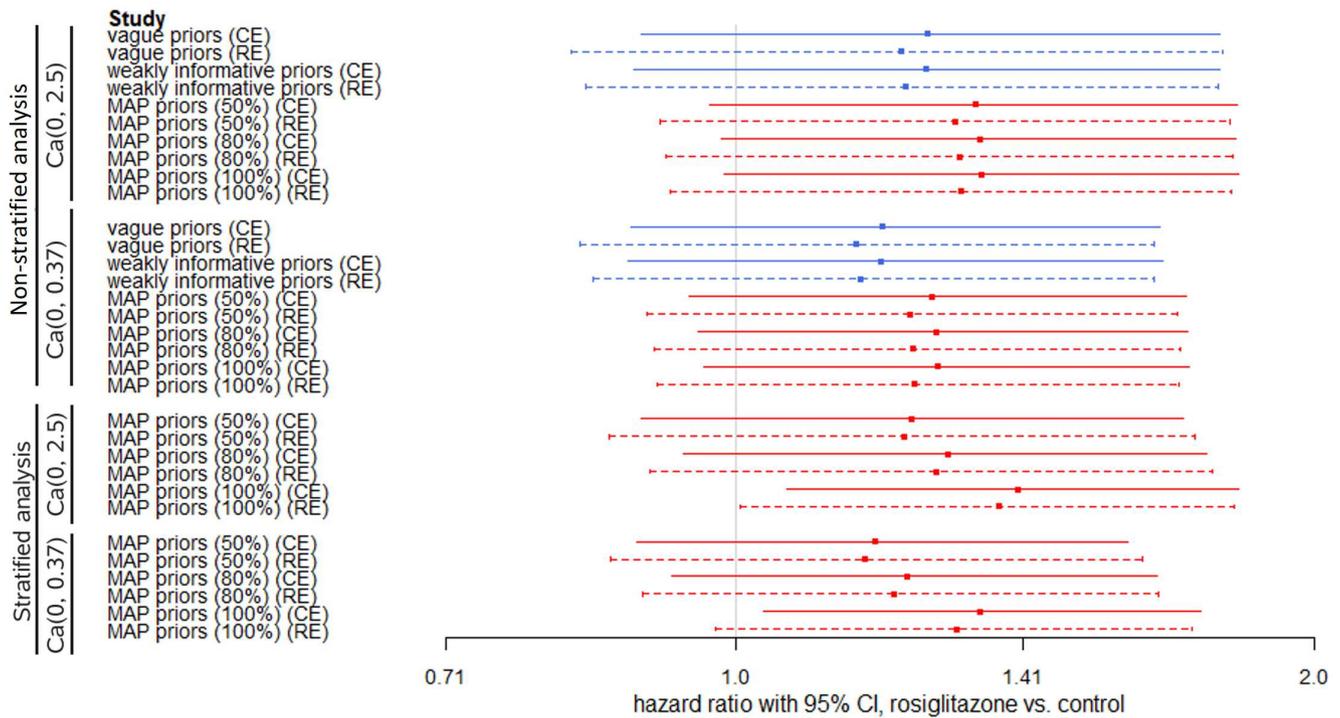}}}
        \caption{Comparison of overall HR~($\varphi$) estimates from CE and RE models in the rosiglitazone example.
        Each line represents the 95\% equal-tailed CI for $\varphi$, the square indicates the point estimate.
        Solid lines show the CIs for CE~models, and dashed lines show RE~models.
        Red and blue lines correspond to models \emph{with} and \emph{without} borrowing from historical trials, respectively.
        Models are first grouped by whether information is borrowed among trials in the main meta-analysis (stratified vs. non-stratified), and then the prior used for~$\varphi$ ($\cauchydistn(0.0, 0.37)$ or $\cauchydistn(0.0, 2.5)$).}
        \label{fig:RosiAppFixedVsRandom}   
    \end{figure}

    For the purpose of comparing common-effect (CE) and random-effects (RE) models, we apply both on the rosiglitazone data set and visualize the point estimate for the log-HR ($\varphi$) along with 95\% equal-tailed credible intervals (CIs) in Figure~\ref{fig:RosiAppFixedVsRandom}. One can see that RE~models yield wider CIs than CE~models in all pairs of comparisons. Because RE~models consider between-trial heterogeneity in addition, the estimation uncertainty is larger than in CE~models. Another finding is that point estimates of the HR in RE~models in this case are generally smaller than in CE~models. Some variation in point estimates is expected, because the weighting of different trials in a meta-analysis will change when moving from a CE~model to a RE~model.\cite{RoeverFriede2021} CE~models yielding larger point estimates mean that trials with larger HR tend to get greater weights, whereas when weights become more even in a RE~model, the impact of these  trials is reduced, and the point estimate, as a weighted average of all trials' HRs, becomes smaller. When checking the rosiglitazone data, we indeed find that trials with large (positive) effect  mostly coincide with large sample size; so this tendency has a reasonable explanation.

\begin{figure}[!htb]
        {\centerline{\includegraphics[scale=0.7]{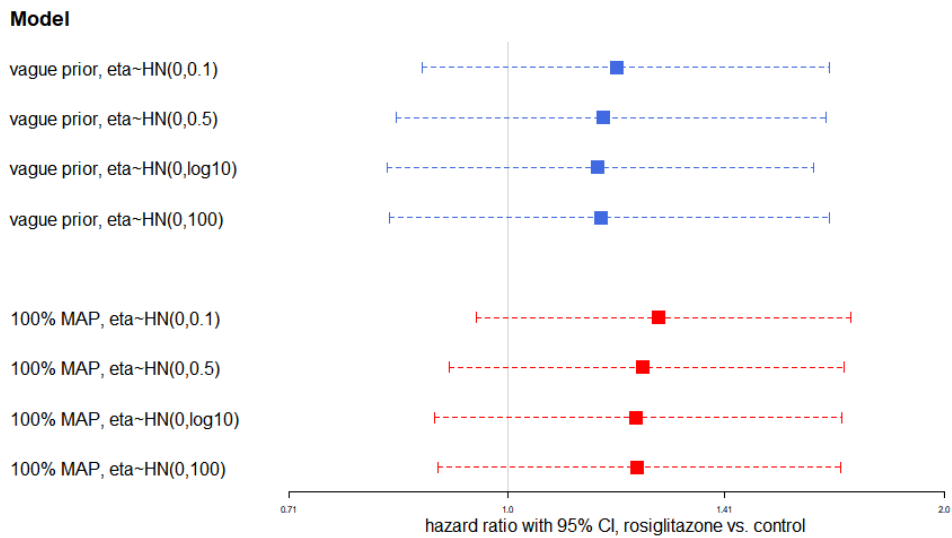}}}
        \caption{Sensitivity analysis for RE~models with varying priors on~$\eta$ in case of a large number of trials in the rosiglitazone example. Blue lines correspond to models without borrowing from historical trials; red lines correspond to models with borrowing from historical trials. Each line represents the 95\% equal-tailed CI for~$\varphi$, the point estimate is indicated by a box on the line. Prior sensitivity (with respect to~$\eta$) is quantified in terms of the $\dcjs$~metric.}
        \label{fig:RosiSensitivityFull}
    \end{figure}

    The proposed Bayesian RE~model naturally takes between-trial heterogeneity into consideration. We want to investigate to what degree variations of prior assumptions on the between-trial heterogeneity~($\eta$) as well as the amount of data considered will impact on the resulting HR estimates. To this end, we perform sensitivity analyses considering either the full data set, or a smaller number of studies. 
    Firstly, we consider all 54~trials in the main meta-analysis. Such a large number of trials constitutes a substantial amount of information on its own. We pick one model \emph{without} borrowing from historical trials and one model \emph{with} borrowing from historical trials as examples, to see the effect of varying priors for the heterogeneity~$\eta$. Table~\ref{tab:RosiSensitivity} shows the details of the selected models, Figure~\ref{fig:RosiSensitivityFull} shows the results. 
    Results are very consistent in both cases with and without inclusion of historical data; only the rather informative $\halfnormaldistn(0.1)$ prior for the heterogeneity yields slightly different results (with estimates closer to those of a CE~model). With this large number of trials, the data seem to dominate as long as the prior is not very informative.

    \begin{table}[ht]
      \caption{Prior specifications used in the sensitivity analyses for the RE~model applied to the rosiglitazone data set. Vague or informative priors are used for the hyperparameters $\nu_1$, $\sigma_1$, $\nu_2$, $\sigma_1$, $q_0$ and~$q_1$ to implement potential borrowing of information (on baseline hazards and fatal event probabilities) from ``historical'' trials. No \emph{robustification} is implemented here, i.e., 100\% weight is given to the MAP priors.
      A weakly informative prior is used for the HR~($\varphi$), as in the CE~model, and a range of increasingly wide priors for the heterogeneity~$\eta$ is investigated.}\label{tab:RosiSensitivity}
      \begin{center}
      \begin{tabular}{rllll}
      \toprule
                   & & \multicolumn{3}{c}{parameters} \\
                     \cmidrule(lr){3-5}
        \multicolumn{2}{l}{model} & $\nu_1$, $\sigma_1$, $\nu_2$, $\sigma_1$, $q_0$, $q_1$ & hazard ratio $\varphi$     & heterogeneity $\eta$ \\
      \midrule
        1 & (no borrowing) & vague priors    & $\cauchydistn(0.0, 0.37)$ & $\halfnormaldistn(0.1)$   \\
        2 &                &                 &                           & $\halfnormaldistn(0.5)$   \\
        3 &                &                 &                           & $\halfnormaldistn\bigl(\log(10)\bigr)$ \\
        4 &                &                 &                           & $\halfnormaldistn(100)$   \\[1ex]
        5 & (borrowing)    & MAP priors (100\%) & $\cauchydistn(0.0, 0.37)$ & $\halfnormaldistn(0.1)$   \\
        6 &                &                 &                           & $\halfnormaldistn(0.5)$   \\
        7 &                &                 &                           & $\halfnormaldistn\bigl(\log(10)\bigr)$ \\
        8 &                &                 &                           & $\halfnormaldistn(100)$   \\
      \bottomrule
      \end{tabular}
      \end{center}
    \end{table}

\begin{figure}[!htb]
        {\centerline{\includegraphics[scale=0.7]{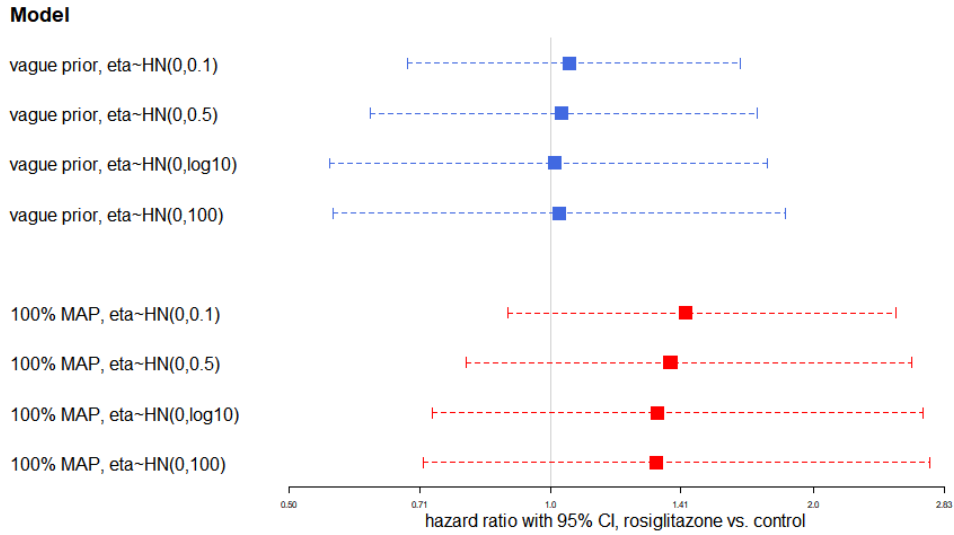}}}
        \caption{Sensitivity analyses for RE~models with varying priors on the heterogeneity~$\eta$ in case of a small number of trials in the rosiglitazone example. Red and blue lines correspond to models \emph{with} and \emph{without} borrowing from historical trials. Each line represents the 95\% equal-tailed CI for~$\varphi$, the point estimate is indicated by a square. Prior sensitivity (with respect to~$\eta$) is quantified in terms of the $\dcjs$~metric.}
        \label{fig:RosiSensitivityFewStudies}
    \end{figure}

    In practice, however, only rather few trials may be available for meta-analysis. We want to also investigate the heterogeneity~($\eta$) of the prior's effect in case of a small number of trials. We select~5 out of the~54 rosiglitazone trials and again implement the models according to Table~\ref{tab:RosiSensitivity}.
    The selected trials are relatively informative ones with larger sample sizes and sizable numbers of events to avoid sparse-data issues. Figure~\ref{fig:RosiSensitivityFewStudies} indicates that, as the heterogeneity ($\eta$) prior becomes more informative, the CIs become narrower and point estimates tend to shift right, both effects indicating a move towards the CE~results. Therefore, in case of small numbers of trials, the data may no longer dominate,  and a sensible and convincing specification of the heterogeneity prior is even more crucial.\cite{RoeverEtAl2021}

    In both scenarios (Figures~\ref{fig:RosiSensitivityFull} and~\ref{fig:RosiSensitivityFewStudies}), prior sensitivity with respect to the heterogeneity~$\eta$ was investigated in terms of the $\dcjs$~metric.\citep{KallioinenEtAl2024} Sensitivity tends to be smallest for a conventional choice of a prior scale of~$0.5$, suggesting that priors favouring more extreme (small or large) heterogeneities may imply a prior/data conflict.
    
  \subsection{Oncology data}\label{sec:OncoAppli}
\begin{figure}[!htb]
        {\centerline{\includegraphics[scale=0.6]{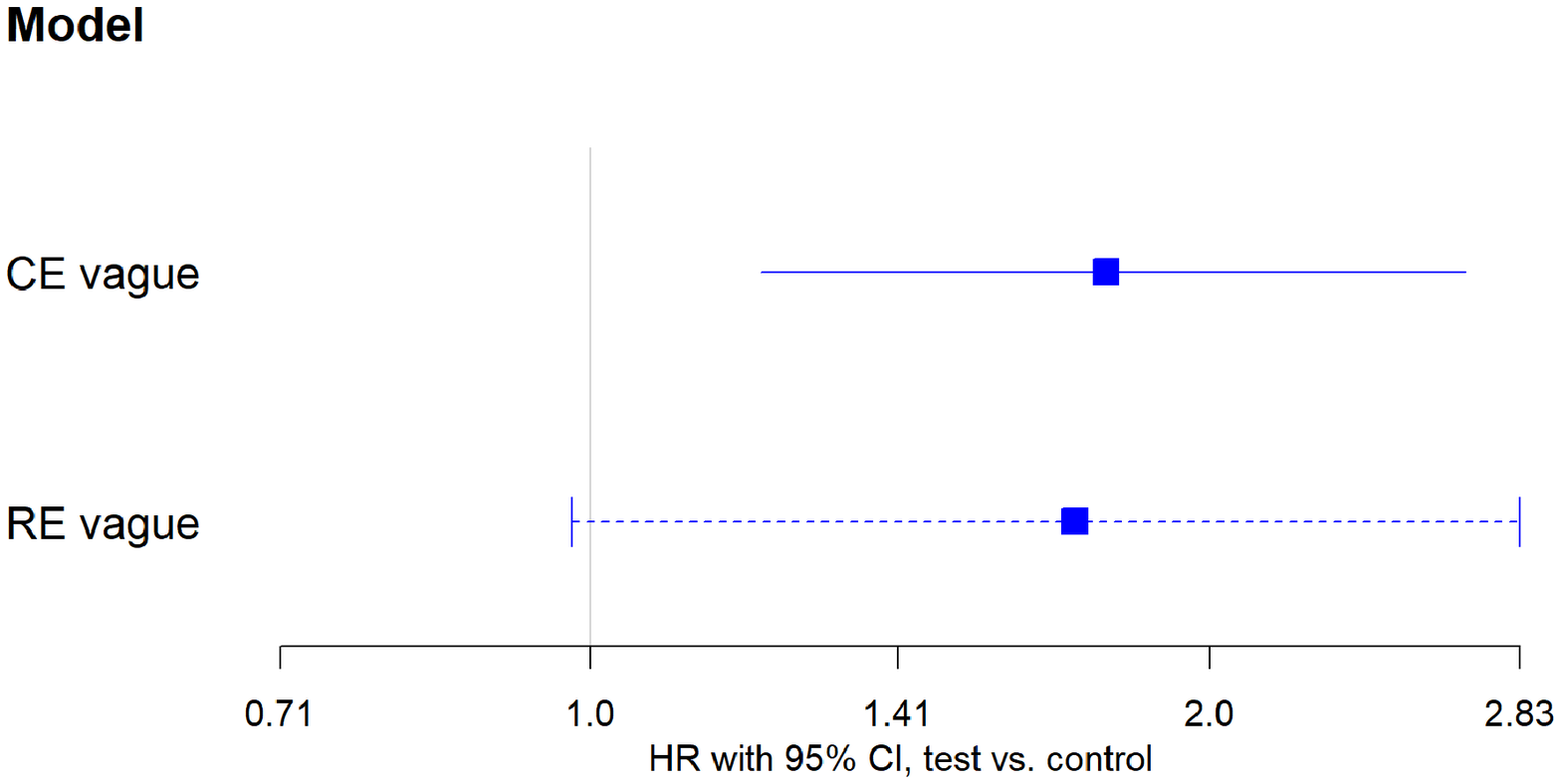}}}
        \caption{Comparison of CE and RE~models in terms of estimation of the overall HR~$\varphi$  in the oncology example. Solid and dashed lines correspond to CE and RE~models for~$\varphi$. Each line represents the 95\% equal-tailed CI for~$\varphi$, the point estimate is indicated by a square.}
        \label{fig:OncoAppFixedVsRandom}
    \end{figure}

    As a second application example, we consider the oncology example data set introduced in Section~\ref{sec:OncoData}. A key difference in the model setup in this application is that priors relate to hyperparameters of the treatment arms (i.e., to $\log(\lambda_{i1})$ instead of $\log(\lambda_{i0})$. This way, $\log(\lambda_{i0})$ has a larger variance than $\log(\lambda_{i1})$, reflecting the heterogeneous control groups observed in the oncology data in Table~\ref{tab:OncoData}. In addition, no historical data were available in this case.
    Estimated HRs are shown in Figure~\ref{fig:OncoAppFixedVsRandom}. Similarly to the rosiglitazone example, the RE~models yield similar point estimates but wider CIs than the CE model. In this case, CE CIs exclude a HR of one, while one is included for the RE-CIs, suggesting qualitatively different conclusions regarding the presence of an effect.

\begin{figure}[!htb]
        {\centerline{\includegraphics[scale=0.6]{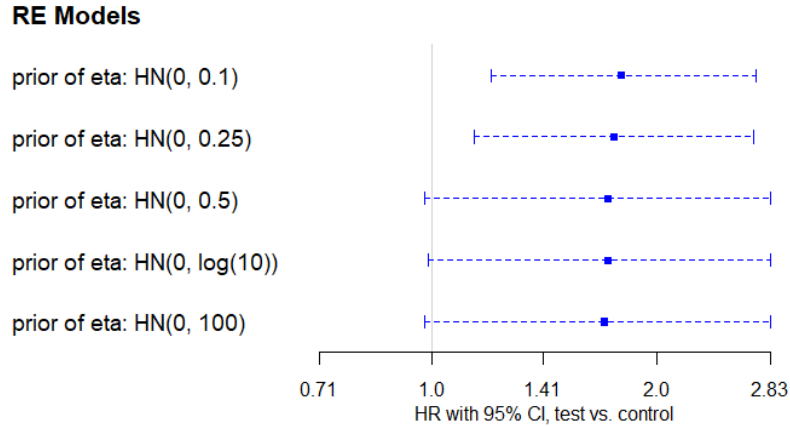}}}
        \caption{Sensitivity analysis for RE~models with varying priors for the heterogeneity~$\eta$ in the oncology example. Each line represents the 95\% equal-tailed CI for the overall HR~$\varphi$, the point estimate is indicated by a square. Prior sensitivity (with respect to~$\eta$) is quantified in terms of the $\dcjs$~metric.}
        \label{fig:OncoSensitivity}
    \end{figure}

    The prior for the heterogeneity~$\eta$ used in the RE~model is~$\halfnormaldistn(0.5)$. We further vary the informativeness of the heterogeneity prior to investigate its impact on the overall HR estimation. Figure~\ref{fig:OncoSensitivity} shows that, in this case with a smaller number of trials, vague or weakly informative priors make little difference, but further increasing informativeness shortens CIs and mitigates estimation uncertainty.
    Again, the $\eta$~prior sensitivity (as quantified through the $\dcjs$~metric) tends to be smallest for a weakly informative $\halfnormaldistn(0.5)$~prior.

  \subsection{Summary}
    In summary, the application on two data sets shows consistent trends.
    The major differences between the applications are the numbers of studies included, as well as the availability of historical trials.
    When varying priors for the heterogeneity~$\eta$ from $\halfnormaldistn(100)$ to $\halfnormaldistn(0.1)$, the CI width for the overall HR~$\varphi$ decreases by~18.4\% in case of the 9~oncology trials, but by only by~5.4\% in case of the 54~Rosiglitazone trials, confirming that the prior plays a more important role in the analysis of the oncology example with fewer studies and no historical data.
    Posterior distributions for the random effects scale (heterogeneity~$\eta$) are illustrated in Figure~\ref{fig:heteroPosterior} (Appendix~\ref{sec:heteroAppendix}), showing that the data only allow provide little constraints on the heterogeneity.

\section{Simulation study}\label{sec:simulStudy}
  \subsection{Aims}
    To investigate the operating characteristics of RE~models in comparison to the CE~model, we performed two simulation studies motivated by the two data examples introduced in Section~\ref{sec:examples}. We will in particular consider coverage probabilities, type~I error and power of the different methods in a limited range of scenarios.
    Additional simulation results investigating departures from the exponential distribution assumption~(\ref{eqn:ExpoAssumption1}) are provided in Appendix~\ref{sec:WeibullAppendix}.

  \subsection{Rosiglitazone scenario}\label{sec:RosiSimul}
    \subsubsection{Setup}\label{sec:sim_rosiglitazone_setup}
      In the first simulation study motivated by the rosiglitazone example data (see Section~\ref{sec:RosiData}), we try to mimic a typical non-oncological drug development process involving 1425~patients comparing a test drug with a control. The main meta-analysis contains two early-stage 12-week trials with 6:1 and 1:1 randomization respectively, three half-year trials with 2:1 randomization, and a single 1-year trial with 1:1 randomization. Sample sizes range from~25 to 300~patients per arm. In addition, there are 12~historical trials with control group data only.\cite{Holzhauer2017} 

      Data are generated based on the model described in Section~\ref{sec:ReModel}.
      We assume the historical trials share the same event rate ($\lambda_{i0}=0.5$) and drop-out rate ($\mu_{i0}=\mu_{i1}=0.5$) with the main meta-analysis. Also, the proportions of fatal events were constant ($q_0=q_1=0.35$).
      We simulate the IPD using parameters shown 
      in Table~\ref{tab:simRosiCover}, 
      and then sum over the patients in each trial-arm to derive the corresponding aggregate data. The major difference between scenarios is in their underlying hazard ratio (HR). We assume either null or positive treatment effects.
      Scenarios 2--4 and 6--8 assume random treatment effects, namely, the trial-specific log-HR~$\varphi_i$ following a $\normaldistn(\varphi, \eta^2)$~distribution (\ref{eqn:RE}).
      When determining the amount of between-trial heterogeneity, we first consider the mean of a $\halfnormaldistn(0.5)$~distribution ($\expect[\eta]=0.40$), as this is considered as a reasonable prior for~$\eta$ when the number of trials is small.\cite{RoeverEtAl2021} Then the value of~$\eta$ is further varied to be half or twice as much, to capture situations with small and large between-trial heterogeneity. In addition, scenarios~1 and~5 assume \emph{no} between-trial heterogeneity ($\eta=0$, homogeneity, CE~assumption). For each scenario, we perform 1000~simulations.
    
      
      For each scenario, the overall HR~$\varphi$ then is estimated using four model variations that result as combinations of the CE~model (see Section~\ref{sec:FeModel}) or its RE extension (see Section~\ref{sec:ReModel}) with or without the consideration of historical control data via a MAP prior. When no historical data are included, \emph{vague} prior settings are used.
      The four modelling approaches are in the following referred to as CE-vague, RE-vague, CE-MAP and RE-MAP\@.

      For the two CE-MAP and RE-MAP models, the meta-analytic-predictive prior is robustified by including a vague mixture component with 50\% weight,\cite{SchmidliEtAl2014}
      and a non-stratified approach is implemented.

    \subsubsection{Results}
      Table~\ref{tab:simRosiCover} shows the coverage rate of 95\% credible intervals (CIs) for the overall HR~$\varphi$ for all four models in each scenario. If we compare the CE models against the RE models, we find that in all scenarios CE models have lower CI coverage rates than RE models, because RE~models account for between-trial heterogeneity in addition to sampling error, leading to wider CIs and increased coverage. For scenarios~2--5 and~7--10, RE models should perform better, as they should be able to account for REs implemented in the data generation. This is supported by the observation from Table~\ref{tab:simRosiCover}; for scenarios~1 and~6, since the data generating mechanism assumes no heterogeneity between trials, the CE model should perform better. Consistently, we indeed find the coverage rate for common effect model in scenarios~1 and~6 to be very close to 95~\%, whereas the CI coverage rates of RE models are greater, suggesting that RE~models are more conservative in these cases.
      The RE~models' coverage then decreases with increasing heterogeneity, and (as expected) may tend towards undercoverage if the heterogeneity setting approaches the upper limit of the a~priori expected range (scenarios~5 and~10).
      
      \begin{table}[ht]
        \caption{Coverage probabilities (in \%) for a nominal 95\% credible interval for the overall HR~($\varphi$) based on the simulations motivated by the rosiglitazone example (as described in Section~\ref{sec:sim_rosiglitazone_setup}).}\label{tab:simRosiCover}
        \begin{center}
          \begin{tabular}{ccllccc}
            \toprule
            \multicolumn{4}{c}{simulation settings} & & \multicolumn{2}{c}{coverage probability (\%)} \\ 
            \cmidrule(lr){1-4} \cmidrule(lr){6-7}
            scenario  & log-HR~$\varphi$ & \multicolumn{2}{l}{heterogeneity~$\eta$} & prior & common effect (CE) & random effects (RE) \\
            \midrule
            1 & 0.00 & 0.00 & (none)     & vague & 94.5 & 98.8 \\
              &      &      &            & MAP   & 96.7 & 99.3 \\[0.5ex]
            2 & 0.00 & 0.10 & (small)     & vague & {92.5} & {98.9} \\
              &      &      &            & MAP   & {94.0} & {99.1} \\[0.5ex]
            3 & 0.00 & 0.20 & (moderate)    & vague & 89.8 & 98.3 \\
              &      &      &            & MAP   & 87.3 & 97.6 \\[0.5ex]
            4 & 0.00 & 0.40 & (substantial) & vague & 72.2 & 97.1 \\
              &      &      &            & MAP   & 68.4 & 94.8 \\[0.5ex]
            5 & 0.00 & 0.80 & (large)    & vague & 45.9 & 97.2 \\
              &      &      &            & MAP   & 43.8 & 92.2 \\[2ex]
            6 & 0.25 & 0.00 & (none)     & vague & 94.9 & 98.6 \\
              &      &      &            & MAP   & 96.9 & 99.0 \\[0.5ex]
            7 & 0.25 & 0.10 & (small)     & vague & {92.8} & {97.3} \\
              &      &      &            & MAP   & {92.3} & {98.7} \\[0.5ex]
            8 & 0.25 & 0.20 & (moderate)    & vague & 89.5 & 98.2 \\
              &      &      &            & MAP   & 86.4 & 97.4 \\[0.5ex]
            9 & 0.25 & 0.40 & (substantial) & vague & 73.9 & 97.6 \\
              &      &      &            & MAP   & 71.4 & 95.4 \\[0.5ex]
           10 & 0.25 & 0.80 & (large)    & vague & 44.1 & 96.6 \\
              &      &      &            & MAP   & 41.0 & 91.0 \\
            \bottomrule
          \end{tabular}
        \end{center}
      \end{table}

      If we further compare the models with and without borrowing from historical trials (using vague priors or MAP priors), we find that in scenarios~2--5 and~7--10, vague models provide higher CI coverage rate than MAP models, while scenarios~1 and~6 show vague models with lower coverage than MAP models. Such trends can be attributed to the difference in point estimates of~$\varphi$ between vague and MAP models. A further investigation shows that for scenarios~2--5 and~7--10, the point estimates of vague and MAP models are very close to each other. Since MAP models have narrower CIs due to additional information borrowed from historical trials, vague models have lower CI coverage than MAP models. However, for scenarios~1 and~6, there is a clear discrepancy between the point estimates of MAP and vague models. The point estimates of MAP models are closer to the true value. Therefore, although MAP models have narrower CIs, they still have higher coverage due to the more precise point estimates.

      Apart from CI coverage, we further investigate the type~I error rate for scenarios under the null hypothesis, and power for scenarios under alternative hypotheses, respectively. Table~\ref{tab:simRosiErrorPower} shows the type~I error rates of scenarios~1--5 which assume~$\varphi$ to be zero (i.e., no adverse effect). For the CE scenario (scenario~1), CE models control the type~I error rate at around~$5\%$, but RE models result in conservative type~I error rates ($>5\%$). For the other scenarios, in contrast to CE models, RE models can better control the type~I error rate, because random effects models tend to have wider CIs and thus lower rejection rates.
      
      \begin{table}[ht]
        \caption{Type~I error rates and power based on the simulations motivated by the rosiglitazone example (as described in Section~\ref{sec:sim_rosiglitazone_setup}).}\label{tab:simRosiErrorPower}
        \begin{center}
          \begin{tabular}{lrcccccccccc}
            \toprule
                  && \multicolumn{5}{c}{type~I error (\%)} & \multicolumn{5}{c}{power (\%)}\\
            \cmidrule(lr){3-7} \cmidrule(lr){8-12}
            model & scenario: & 1 & 2 & 3 & 4 & 5 & 6 & 7 & 8 & 9 & 10\\
            \midrule
            \multicolumn{2}{l}{CE-vague} & \phantom{0}5.5 & {7.5} & 10.2 & 27.8 & 54.1
                                         & 51.4 & {51.8} & 55.4 & 61.5 & 73.0 \\
            \multicolumn{2}{l}{CE-MAP}   & \phantom{0}3.3 & {6.0} & 12.7 & 31.6 & 56.2
                                         & 79.2 & {75.6} & 73.0 & 68.6 & 74.8 \\
            \multicolumn{2}{l}{RE-vague} & \phantom{0}1.2 & {1.1} & \phantom{0}1.7 & \phantom{0}2.9 & \phantom{0}2.8
                                         & 29.0 & {24.5} & 22.6 & 12.4 & \phantom{0}7.0 \\
            \multicolumn{2}{l}{RE-MAP}   & \phantom{0}0.7 & {0.9} & \phantom{0}2.4 & \phantom{0}5.2 & \phantom{0}7.8
                                         & 47.5 & {45.2} & 35.7 & 22.0 & 16.6 \\
            \bottomrule
          \end{tabular}
        \end{center}
      \end{table}

      Table~\ref{tab:simRosiErrorPower} also shows the power for scenarios~6--10 which assume~$\varphi$ to be different from zero. We focus on the models successfully controlling the type~I error rate in the same table. For the RE models in Table~\ref{tab:simRosiErrorPower}, MAP models have greater power than vague models, because MAP models with additional historical information yield narrower CIs and thus higher rejection rates. Since the underlying data generating mechanism assumes a CE in scenario~6, CE~models outperform RE~models in terms of power, as expected.  
      The accuracy of point estimates for the log-HR is summarized in Table~\ref{tab:simRosiPointEsti} (Appendix~\ref{sec:PointEstiAppendix}).

      To summarize, in the presence of between-trial heterogeneity, RE~models tend to perform better than CE~models in terms of CI~coverage and type~I error control. In case of homogeneity, RE~models are too conservative, leading to too high CI~coverage, too low type~I error rate and low power. The situation could be mitigated by choosing a smaller scale for the prior of~$\eta$ in such cases. For example, for scenarios~1 and~6, if we set the prior of~$\eta$ to be $\halfnormaldistn(0.25)$ instead of $\halfnormaldistn(100)$ for vague models and $\halfnormaldistn(0.5)$ for MAP models, the type~I error rate of RE models for scenario~1 increases from around 1\% to around 2\%, while the power of RE models for scenario~5 increases from 29.0\% to 38.5\% for the vague model and from 47.5\% to 58.4\% for the MAP model, respectively.

  \subsection{Oncology scenario}
    \subsubsection{Setup}\label{sec:simOncoSetup}
      Since oncology trials have many unique features, we considered it worthwhile to separately conduct a simulation study to investigate the behavior of the proposed models under such circumstances. The simulation design is motivated by the oncology data example introduced in Section~\ref{sec:OncoData}. We fit the RE model using vague priors to the data and designed simulation scenarios listed 
      in Table~\ref{tab:simOncoCover}
      based on the derived estimates. The major difference among scenarios still lies in the treatment effects. Following the same structure as the rosiglitazone simulation, we make the treatment effects be random for most scenarios and fixed for the scenarios~1 and~5. Either a null or positive treatment effect is assumed for the value of the HR~$\varphi$. 
      Between-trial heterogeneity varies from 0.0 (homogeneity), over~0.1 (small), 0.2 (moderate) and~0.4 (substantial) to~0.8 (large).
      

      A difference in contrast to the rosiglitazone simulation needs to be pointed out; in Section~\ref{sec:sim_rosiglitazone_setup}, we specify a fixed $\log(\lambda_{i0})$ and a random $\varphi_i$. Hence, $\log(\lambda_{i1})$ is also random because $\log(\lambda_{i1})=\log(\lambda_{i0})+\varphi_i$.
      However, in the oncology simulation, we set both $\log(\lambda_{i1})$ and $\varphi_i$ to be random. As a result, $\log(\lambda_{i0})$ is also random and is derived by $\log(\lambda_{i0})=\log(\lambda_{i1})-\varphi_i$. 
      So $\log(\lambda_{i0})$ has larger variance than $\log(\lambda_{i1})$, 
      because $\var\bigl(\log(\lambda_{i0})\bigr)=\var\bigl(\log(\lambda_{i1})\bigr)+\var(\varphi_i)$. 
      This is in line with our observation in the oncology data example that treatment arms of all trials containing the same therapy of interest are similar, while control arms are heterogeneous in different trials. 
      The number of trials, arm-level sample size and maximum duration are set roughly according to the oncology data (see Section~\ref{sec:OncoData}).
      For the treatment group's log-hazard rate, we assume $\log(\lambda_{i1})\sim\normaldistn\bigl(\nu_1=\log(0.02), \sigma_1^2=1.2^2\bigr)$.
      The drop-out hazard rates are the same and constant in both groups ($\mu_{i0}=\mu_{i1}=0.5$, i.e., $\nu_2=\nu_3=\log(0.5)$ and $\sigma_2=\sigma_3=0$), 
      as are the probabilities for fatal events ($q_0=q_1=0.01$).
      We fit CE and RE models \emph{without} borrowing from historical trials to each simulated data set, since no historical data are available.

    \subsubsection{Results}
      Table~\ref{tab:simOncoCover} shows the coverage rate of 95\% equal-tailed CIs of both CE and RE models in each scenario.
      For all scenarios, CE models have lower CI coverage than RE models, as expected, since RE models additionally consider between-trial heterogeneity and thus have wider CIs. 
      For RE scenarios (scenarios~2--5 and 7--10), RE models should give CI coverage closer to 95\% than CE models. 
      But for CE scenarios (scenarios~1 and~6), the CI coverage of CE models should be closer to 95\% than RE models. Results in Table~\ref{tab:simOncoCover} match our expectation well. 
      Focusing on scenarios~1--5 in Table~\ref{tab:simOncoCover}, we observe a decrease in CI coverage as the heterogeneity~$\eta$ increases, for both CE and RE models. The same trend can be observed for scenarios~6--10. 
      The obvious explanation is that when the true value of~$\eta$ is larger, CE models tend to underestimate the total variation more, thus the resulting CI is much narrower, and coverage decreases. For RE models, scenarios~4 and~8 have CI coverage rate quite close to 95\% because the prior for~$\eta$ ($\halfnormaldistn(0.5)$) is consistent with its true value~($\eta=0.4$). For scenarios~3 and~8, the prior expectation of~$\eta$ is larger than its true value, thus the between-trial heterogeneity is overestimated, resulting in too wide CIs and an increased (conservative) coverage rate. For scenarios~5 and~10, the prior expectation on~$\eta$ is lower than its true value, so the between-trial heterogeneity gets underestimated, leading to too narrow CIs and thus too low coverage rate.
      
      \begin{table}[ht]
        \caption{Coverage probabilities (in \%) for a nominal 95\% credible interval for the overall HR~($\varphi$) based on the simulations motivated by the oncology example (as described in Section~\ref{sec:simOncoSetup}).}\label{tab:simOncoCover}
        \begin{center}
          \begin{tabular}{ccllccc}
            \toprule
            \multicolumn{4}{c}{simulation settings} &  & \multicolumn{2}{c}{coverage probability (\%)}\\
            \cmidrule(lr){1-4} \cmidrule(lr){6-7}
            scenario & log-HR~$\varphi$ & \multicolumn{2}{l}{heterogeneity~$\eta$} & prior & common effect (CE) & random effects (RE) \\
            \midrule
            1  & 0.00 & 0.00 & (none)     &  vague  &  95.9  &  98.5  \\
            2  & 0.00 & 0.10 & (small)     &  vague  &  {93.2}  &  {98.0}  \\
            3  & 0.00 & 0.20 & (moderate)    &  vague  &  86.1  &  96.8  \\
            4  & 0.00 & 0.40 & (substantial) &  vague  &  68.8  &  94.3  \\
            5  & 0.00 & 0.80 & (large)    &  vague  &  40.7  &  89.9  \\[1ex]
            6  & 0.50 & 0.00 & (none)     &  vague  &  94.7  &  98.4  \\
            7  & 0.50 & 0.10 & (small)     &  vague  &  {92.7}  &  {97.7}  \\
            8  & 0.50 & 0.20 & (moderate)    &  vague  &  88.6  &  98.3  \\
            9  & 0.50 & 0.40 & (substantial) &  vague  &  71.5  &  95.6  \\
            10  & 0.50 & 0.80 & (large)    &  vague  &  43.6  &  91.1  \\
            \bottomrule
          \end{tabular}
        \end{center}
      \end{table}

      As in the previous simulation (Section~\ref{sec:RosiSimul}), we also investigate other statistical properties such as the type~I error and power. 
      Table~\ref{tab:simOncoErrorPower} shows the type~I error rates of no-treatment-effect scenarios~1--5. For scenario~1, which assumes no between-trial heterogeneity, the CE~model can control the type~I error rate at below~5\%, while the RE~model results in too conservative type~I error rate. For all five scenarios, RE~models have lower type~I error rates than CE~models. We focus on cases where type~I error is controlled in Table~\ref{tab:simOncoErrorPower} and consider their power. Under a CE~scenario (scenario~6), the CE~model achieves high power (above 90\%), while the RE~model attains lower power due to their larger CI~width. For RE~models, as between-trial heterogeneity increases from scenario~6 to scenario~7, 8, 9 and~10, the power decreases. This is because higher heterogeneity leads to lower signal-to-noise ratio. As a result, the signal is more difficult to identify, and hence correct rejection is more difficult.
      The accuracy of point estimates for the log-HR is summarized in Table~\ref{tab:simOncoPointEsti} (Appendix~\ref{sec:PointEstiAppendix}).

      \begin{table}[ht]
        \caption{Type~I error rates and power based on the simulations motivated by the oncology example (as described in Section~\ref{sec:simOncoSetup}).}\label{tab:simOncoErrorPower}
        \begin{center}
          \begin{tabular}{lrcccccccccc}
            \toprule
                  & & \multicolumn{5}{c}{type~I error (\%)} & \multicolumn{5}{c}{power (\%)} \\
            \cmidrule(lr){3-7} \cmidrule(lr){8-12}
            model & scenario: &  1 & 2 & 3 & 4 & 5 & 6 & 7 & 8 & 9 & 10 \\
            \midrule
            \multicolumn{2}{l}{CE-vague} & 4.1 & {6.8} & 13.9 & 31.2 & 59.3
                                         & 91.1 & {85.8} & 86.0 & 74.1 & 60.7 \\
            \multicolumn{2}{l}{RE-vague} & 1.5 & {2.0} & \phantom{0}3.2 & \phantom{0}5.7 & 10.1
                                         & 79.0 & {75.1} & 72.5 & 57.5 & 34.8 \\
            \bottomrule
          \end{tabular}
        \end{center}
      \end{table}

      In summary, analogous trends are observed in the oncology simulation as in the rosiglitazone simulation. The only difference we find is in the absolute magnitude of power. Comparing Tables~\ref{tab:simRosiErrorPower} and~\ref{tab:simOncoErrorPower}, the power for the oncology simulation is much higher than in the rosiglitazone simulation. This may most likely be attributed to the larger effect~($\varphi$) in the oncology simulations,
      leading to easier rejection and thus higher power in the oncology simulation.

\section{Discussion}\label{sec:discussion}
  In evidence synthesis and meta-analysis contexts, the one-to-one correspondence of results from separate experiments or studies is often questionable, due to differences in designs, follow-up times, or definition of AEs. In addition, differences in capturing, adjudicating and reporting clinical events can increase between-study heterogeneity.
  The anticipation and consideration of (potential) \emph{heterogeneity} between data sources hence is commonly advocated.\cite{Higgins2008,CochraneHandbook} On the technical side, implementation of analysis models accounting for heterogeneity usually leads to more cautious or conservative inferences.\cite{BorensteinEtAl2010,KontopantelisSpringateReeves2013}
  We have extended Holzhauer's proposed meta-analysis model for synthesizing adverse event counts\cite{Holzhauer2017} to incorporate a random-effect (RE) component in order to safeguard against overly optimistic or overconfident conclusions. 
  When Holzhauer presented the original model, it was applied in an example case\cite{NissenWolski2007a} that had already triggered some controversy,\cite{RueckerSchumacher2008,FriedrichEtAl2009,NissenWolski2010} and he was able to advance the discussion by leveraging additional external evidence via an informative MAP~prior.\cite{SchmidliEtAl2014}
  We were able to demonstrate that the consideration of heterogeneity and use of an RE~model also makes a difference here; using the rosiglitazone example as well as an additional application in oncology, we could show the impact of model specification details on the conclusions for these particular applications as well as in some limited Monte Carlo studies. 
  Use of meta-analysis methods as well as consideration of heterogeneity between data sources allows to broaden the evidence base for decision making beyond the limits of single clinical trials that are usually powered with a focus on certain specific efficacy endpoints.
  Consideration of additional data will allow for more definite statements about harms and benefits, while allowing for exploratory investigations into wider ranges of questions. In the examples, very different disease areas were considered. Depending on the disease, only very specific AEs of interst qualify for such summaries as otherwise the noise from the anticipated events (i.e. those seen in the population of interest regardless of investigational treatment) may be too strong.

  Note that model specification in terms of a random ``control'' response (here: $\lambda_{i0}$) and a random treatment effect ($\varphi_i$) implies greater variability in the treatment responses compared to the control arms. In general, this is an often neglected feature, in the oncology example (Section~\ref{sec:OncoAppli}), this was explicitly exploited in order to allow for greater variability for the (rather heterogeneous) control arms. Alternatively, one could specify a ``symmetric'' model variation as in Jackson \emph{et~al.} (2018),\cite{JacksonEtAl2018}
  which may be preferrable in some applications.
  Another common issue in oncology applications is the use of active comparators, so that the extension to network meta-analysis methods may be an option; as long as only two-armed trials are included, this may be accommodated in a rather simple meta-regression model.\citep{RoeverFriede2023}
  Another aspect in which a more realistic model specification might be worth considering is a possible extension to a competing-risks model.\citep{AllignolEtAl2016}

  The approach for adverse-event analysis considered here largely relies on the availability of relevant data and their reporting to a common standard.
  Unfortunately this is often not the case, hindering the assessment of a treatment's risks and benefits.
  The analysis of clinical events and the comparison across trials would be much easier if the relevant data (such as HRs) were more commonly available; the establishment of consistent reporting standards would be very helpful in this context.\citep{RufibachEtAl2024}
  
  The borrowing mechanism implemented in order to include external, historical data may be extended to potentially address additional requirements. 
  \emph{Robustification} was accomplished here via a mixture prior,\citep{SchmidliEtAl2014} whose design features (vague component, weights) need to be chosen carefully in order to avoid unintended behaviour. 
  Alternatively, this may be addressed by explicitly implementing a more sophisticated joint likelihood reflecting the exact assumptions on the relationship between external and current data in terms of a \emph{bias allowance} model.\citep{WeltonEtAl,Holzhauer2020}

  In the simulations and example applications we used vague prior specifications (such as a variance of~$100^2$ for location parameter priors). In the present context, these were meant to reflect conservative defaults, while in any specific practical application, there may be some prior information providing reasonable constraints that may be worth considering in the analysis.\citep{SpiegelhalterEtAl}
  For location parameters, it may be most sensible to specify priors as ``neutral'' (e.g., with zero mean for log-HRs) and vague, where the magnitude of the prior variance may be motivated by the range of effect sizes considered plausible, or relative to the variance that would correspond to the information conveyed by a single observation.\citep{KassWasserman1995}
  Variance (heterogeneity) parameters may be specified based on general considerations of plausible amounts of variation between studies,\citep{RoeverEtAl2021} or sometimes also with reference to relevant empirical evidence.\citep{RoeverEtAl2023,LilienthalEtAl2024}

  When considering time-to-event data, an exponential model for waiting times may be an obvious starting point, while the validity of such an assumption may sometimes be questionable. In our simulations, we also considered scenarios with violations of this assumption, which are shown in Appendix~\ref{sec:WeibullAppendix}. As expected, operating characteristics deteriorate when the modelling assumption is not met; in practice, in case specific assumptions on the time-to-event model may be hard to justify, implementation of more flexible families (like Weibull, Lomax or piecewise exponential distributions) may be desirable.
  While in the applications presented here we used an implementation using \textsf{Stan},\citep{R:rstan} analyses may also be formulated using the \texttt{brms} \textsf{R}~package, which may also simplify extensions of the models considered here.\citep{Buerkner2017}



\ack
\section*{Acknowledgment}
The GWDG High Performance Computing cluster (\url{https://gwdg.de/en/hpc/}) was used to perform the computationally demanding simulations. 

\section*{Conflicts of interest}
Anja Loos is an employee of Merck KgaA. The other authors have declared no conflict of interest.

\section*{Data availability}
The rosiglitazone example data are available in the supplementary material of Holzhauer (2017),\cite{Holzhauer2017} and the oncology data are listed in Table~\ref{tab:OncoData}. The \textsf{R}~code to reproduce calculations is available on GitHub (\url{https://github.com/wuqiongruby/RE\_MA\_AD}).



\bibliographystyle{wileyNJD-AMA}
\bibliography{literature}

\clearpage

\begin{appendix}
\section{Heterogeneity posteriors}\label{sec:heteroAppendix}
  \begin{figure}
    \centering\makebox{\includegraphics[width=0.40\linewidth]{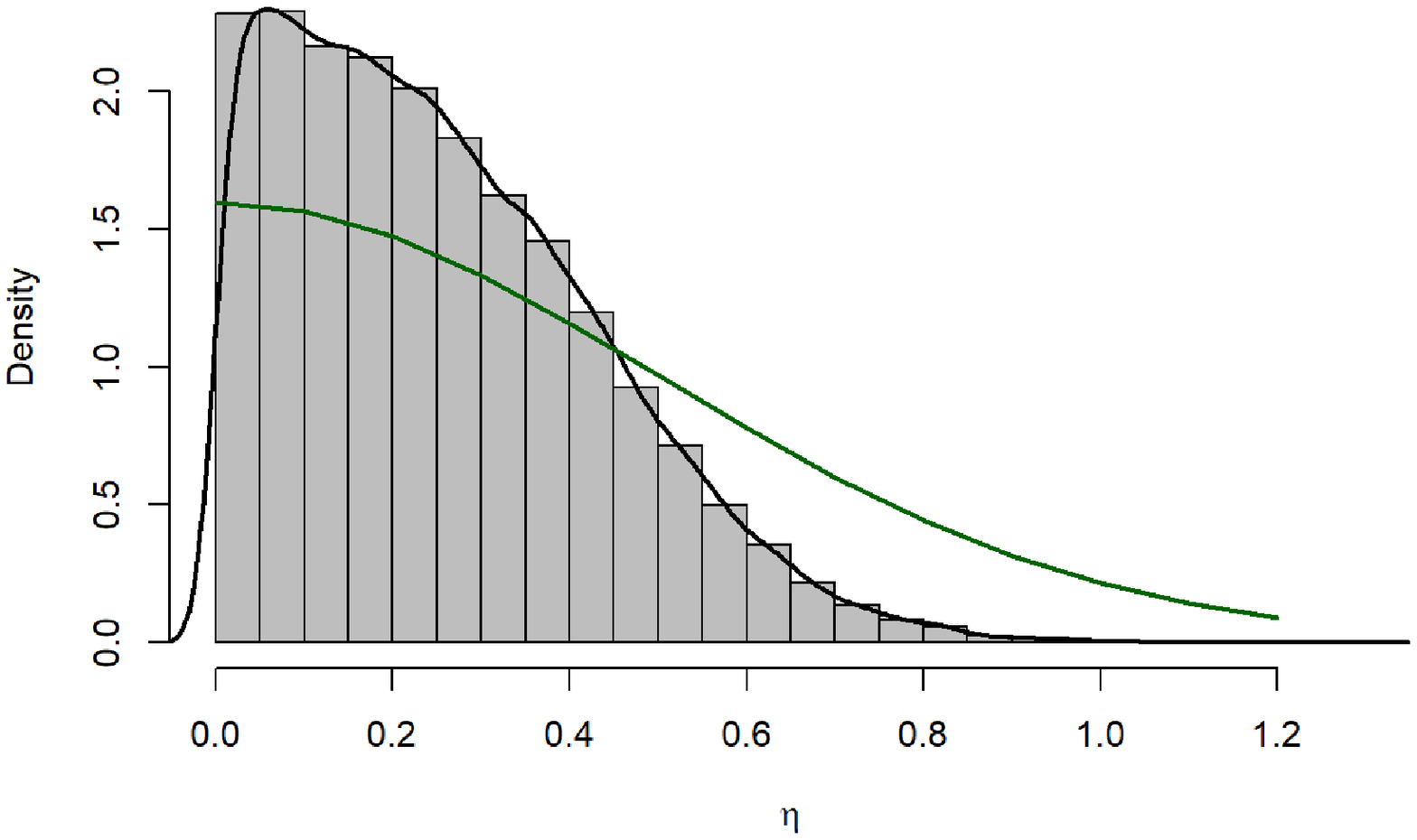} \includegraphics[width=0.40\linewidth]{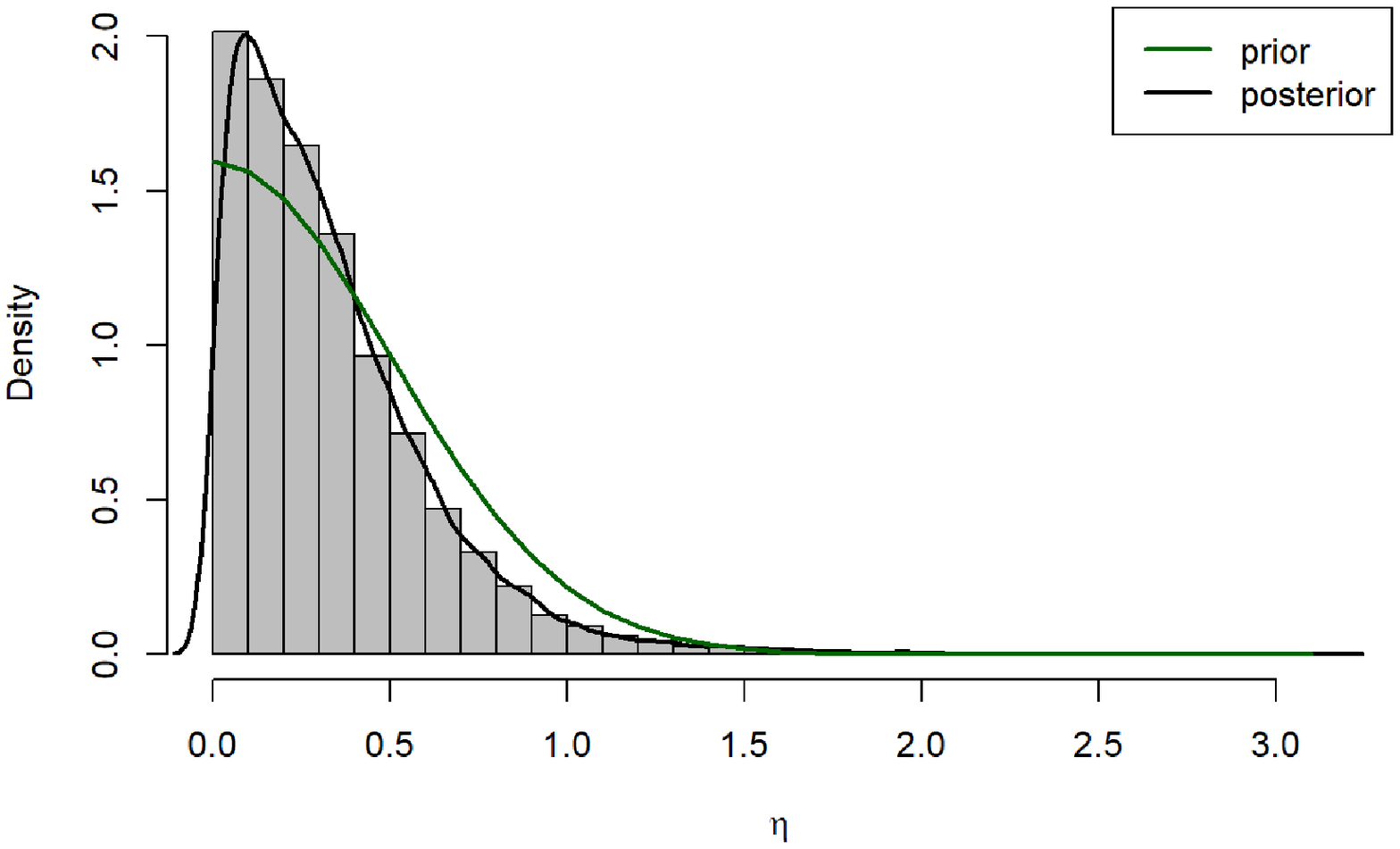}}
    \caption{\label{fig:heteroPosterior} Illustration of heterogeneity posteriors along with the prior distributions for the rosiglitazone and oncology examples discussed in Section~\ref{sec:exampleAppli}.}
  \end{figure}

Figure~\ref{fig:heteroPosterior} illustrates the posteriors for the heterogeneity parameters~($\eta$) for two selected analyses of the rosiglitazone and oncology data sets from Section~\ref{sec:exampleAppli} (the non-stratified analysis of the rosiglitazione data set, using a weakly informative $\cauchydistn(0.0,\,0.37)$ prior for the effect, and the analysis using a vague prior for the effect for the oncology data). In both cases, a $\halfnormaldistn(0.5)$ prior was specified for~$\eta$.

\section{Point estimation}\label{sec:PointEstiAppendix}
  Table~\ref{tab:simRosiPointEsti} illustrates the behaviour of point estimates (log-HRs, in terms of their means and standard deviations) for the simulations motivated by the rosiglitazone example (see Section~\ref{sec:sim_rosiglitazone_setup}.
      \begin{table}[ht]
        \caption{Performance of point estimates (in terms mean and standard deviations) of log-HRs for the simulation settings as described in Section~\ref{sec:sim_rosiglitazone_setup}.}\label{tab:simRosiPointEsti}
        \begin{center}
          \begin{tabular}{ccllccc}
            \toprule
            \multicolumn{4}{c}{simulation settings} & & \multicolumn{2}{c}{posterior mean log-HR: mean (st.dev.)} \\ 
            \cmidrule(lr){1-4} \cmidrule(lr){6-7}
            scenario  & log-HR~$\varphi$ & \multicolumn{2}{l}{heterogeneity~$\eta$} & prior & common effect (CE) & random effects (RE) \\
            \midrule
            1 & 0.00 & 0.00 & (none)     & vague & $0.007$ ($0.133$) & $-0.001$ ($0.134$) \\
              &      &      &            & MAP   & $0.002$ ($0.090$) & $-0.006$ ($0.092$) \\[0.5ex]
            2 & 0.00 & 0.10 & (small)     & vague & $0.002$ ($0.145$) & $-0.007$ ($0.145$) \\
              &      &      &            & MAP   & $-0.005$ ($0.100$) & $-0.015$ ($0.101$) \\[0.5ex]
            3 & 0.00 & 0.20 & (moderate)    & vague & $0.020$ ($0.159$) & $0.003$ ($0.159$) \\
              &      &      &            & MAP   & $0.009$ ($0.126$) & $-0.008$ ($0.124$) \\[0.5ex]
            4 & 0.00 & 0.40 & (substantial) & vague & $0.065$ ($0.222$) & $0.016$ ($0.207$) \\
              &      &      &            & MAP   & $0.056$ ($0.201$) & $0.004$ ($0.185$) \\[0.5ex]
            5 & 0.00 & 0.80 & (large)    & vague & $0.198$ ($0.390$) & $0.001$ ($0.345$) \\
              &      &      &            & MAP   & $0.193$ ($0.385$) & $0.008$ ($0.334$) \\[2ex]
            6 & 0.25 & 0.00 & (none)     & vague & $0.256$ ($0.130$) & $0.249$ ($0.131$) \\
              &      &      &            & MAP   & $0.246$ ($0.085$) & $0.239$ ($0.086$) \\[0.5ex]
            7 & 0.25 & 0.10 & (small)     & vague & $0.250$ ($0.143$) & $0.243$ ($0.144$) \\
              &      &      &            & MAP   & $0.247$ ($0.098$) & $0.239$ ($0.099$) \\[0.5ex]
            8 & 0.25 & 0.20 & (moderate)    & vague & $0.268$ ($0.154$) & $0.253$ ($0.152$) \\
              &      &      &            & MAP   & $0.262$ ($0.122$) & $0.247$ ($0.120$) \\[0.5ex]
            9 & 0.25 & 0.40 & (substantial) & vague & $0.310$ ($0.220$) & $0.254$ ($0.207$) \\
              &      &      &            & MAP   & $0.298$ ($0.201$) & $0.244$ ($0.186$) \\[0.5ex]
           10 & 0.25 & 0.80 & (large)    & vague & $0.441$ ($0.409$) & $0.239$ ($0.364$) \\
              &      &      &            & MAP   & $0.435$ ($0.407$) & $0.243$ ($0.351$) \\
            \bottomrule
          \end{tabular}
        \end{center}
      \end{table}
  Similarly, Table~\ref{tab:simOncoPointEsti} eveluates point estimation in the simulation settings motivated by the oncology example (see Section~\ref{sec:simOncoSetup}).
      \begin{table}[ht]
        \caption{Performance of point estimates (in terms mean and standard deviations) of log-HRs for the simulation settings as described in Section~\ref{sec:simOncoSetup}.}\label{tab:simOncoPointEsti}
        \begin{center}
          \begin{tabular}{ccllccc}
            \toprule
            \multicolumn{4}{c}{simulation settings} & & \multicolumn{2}{c}{posterior mean log-HR: mean (st.dev.)} \\ 
            \cmidrule(lr){1-4} \cmidrule(lr){6-7}
            scenario  & log-HR~$\varphi$ & \multicolumn{2}{l}{heterogeneity~$\eta$} & prior & common effect (CE) & random effects (RE) \\
            \midrule
            1 & 0.00 & 0.00 & (none)     & vague & $0.011$ ($0.131$) & $0.028$ ($0.138$) \\
            2 & 0.00 & 0.10 & (small)     & vague & $0.006$ ($0.159$) & $0.028$ ($0.164$) \\
            3 & 0.00 & 0.20 & (moderate)    & vague & $-0.019$ ($0.167$) & $0.005$ ($0.166$) \\
            4 & 0.00 & 0.40 & (substantial) & vague & $-0.054$ ($0.237$) & $-0.005$ ($0.215$) \\
            5 & 0.00 & 0.80 & (large)    & vague & $-0.201$ ($0.421$) & $-0.032$ ($0.335$) \\[1ex]
            6 & 0.50 & 0.00 & (none)     & vague & $0.505$ ($0.158$) & $0.531$ ($0.172$) \\
            7 & 0.50 & 0.10 & (small)     & vague & $0.503$ ($0.172$) & $0.529$ ($0.179$) \\
            8 & 0.50 & 0.20 & (moderate)    & vague & $0.489$ ($0.180$) & $0.516$ ($0.180$) \\
            9 & 0.50 & 0.40 & (substantial) & vague & $0.438$ ($0.245$) & $0.491$ ($0.228$) \\
           10 & 0.50 & 0.80 & (large)    & vague & $0.289$ ($0.427$) & $0.460$ ($0.337$) \\
            \bottomrule
          \end{tabular}
        \end{center}
      \end{table}

\section{Non-exponential time-to-event}\label{sec:WeibullAppendix}
In order to investigate departures from the assumption of exponentially distributed waiting times for events (equation~(\ref{eqn:ExpoAssumption1})), we initiated limited simulations based on Weibull distributed times as variations of the simulations described in Sections~\ref{sec:sim_rosiglitazone_setup} and~\ref{sec:simOncoSetup}.
As expected, with violated assumptions the operating characteristics tend to worsen; details on coverage probabilities, type-I error and power are provieded in Tables~\ref{tab:simRosiWeibull} and~\ref{tab:simOncoWeibull}.
It should also be noted that in our implementation, use of the Weibull distribution for data generation led to increased rates of non-convergence for the MCMC, suggesting that longer burn-in and sampling times may be appropriate.

      \begin{table}[ht]
        \caption{Coverage probabilities, type-I error and power for variations of the simulation scenarios described in Section~\ref{sec:sim_rosiglitazone_setup}. Heterogeneity is at $\eta=0.2$ throughout, and Weibull-distributed times-to-event were simulated with several settings for the shape parameter.}\label{tab:simRosiWeibull}
        \begin{center}
          \begin{tabular}{cccccccccc}
            \toprule
            \multicolumn{3}{c}{simulation settings} & & \multicolumn{2}{c}{coverage (\%)} & \multicolumn{2}{c}{type-I error (\%)} & \multicolumn{2}{c}{power (\%)}\\ 
            \cmidrule(lr){1-3} \cmidrule(lr){5-6} \cmidrule(lr){7-8} \cmidrule(lr){9-10}
            scenario  & log-HR~$\varphi$ & Weibull shape & prior & CE & RE & CE & RE & CE & RE \\
            \midrule
            3 & 0.00 & 1.5 & vague & 66.4 & 97.6 & 33.6 &  2.4 &  &  \\
              &      &     & MAP   & 64.7 & 90.6 & 35.3 &  9.4 &  &  \\[0.5ex]
            3 & 0.00 & 2.0 & vague & 64.9 & 97.6 & 35.1 &  2.4 &  &  \\
              &      &     & MAP   & 61.2 & 89.4 & 38.8 & 10.6 &  &  \\[0.5ex]
            3 & 0.00 & 2.5 & vague & 63.2 & 98.5 & 36.8 &  1.5 &  &  \\
              &      &     & MAP   & 61.3 & 87.8 & 38.7 & 12.2 &  &  \\[0.5ex]
            8 & 0.25 & 1.5 & vague & 57.2 & 96.8 &      &      & 68.1 & 10.4 \\
              &      &     & MAP   & 56.5 & 93.6 &      &      & 66.0 & 13.6 \\[0.5ex]
            8 & 0.25 & 2.0 & vague & 47.0 & 96.2 &      &      & 69.9 & 7.7 \\
              &      &     & MAP   & 48.9 & 91.4 &      &      & 66.0 & 16.5 \\[0.5ex]
            8 & 0.25 & 2.5 & vague & 40.8 & 97.4 &      &      & 72.4 & 6.6 \\
              &      &     & MAP   & 42.8 & 86.1 &      &      & 68.8 & 23.2 \\
            \bottomrule
          \end{tabular}
        \end{center}
      \end{table}

      \begin{table}[ht]
        \caption{Coverage probabilities, type-I error and power for variations of the simulation setups described in Section~\ref{sec:simOncoSetup}. Heterogeneity is at~$\eta=0.2$ throughout, and times-to-event are simulated based on a Weibull distribution with shape~$1.5$.}\label{tab:simOncoWeibull}
        \begin{center}
          \begin{tabular}{cccccccccc}
            \toprule
            \multicolumn{3}{c}{simulation settings} & & \multicolumn{2}{c}{coverage (\%)} & \multicolumn{2}{c}{type-I error (\%)} & \multicolumn{2}{c}{power (\%)}\\ 
            \cmidrule(lr){1-3} \cmidrule(lr){5-6} \cmidrule(lr){7-8} \cmidrule(lr){9-10}
            scenario  & log-HR~$\varphi$ & Weibull shape & prior & CE & RE & CE & RE & CE & RE \\
            \midrule
            3 & 0.0 & 1.5 & vague & 70.8 & 93.4 & 29.2 &  6.6 &  &  \\
            8 & 0.5 & 1.5 & vague & 67.9 & 85.1 &      &      & 66.1 & 55.6 \\
            \bottomrule
          \end{tabular}
        \end{center}
      \end{table}

\end{appendix}

\end{document}